\documentclass[11pt,onecolumn,amssymb,nofootinbib]{revtex4}
\usepackage{amsmath, amsthm, amscd, amssymb}
\usepackage{bm}
\usepackage{bbm}

\begin{document}

\title{\bf Coleman-Weinberg symmetry breaking in $SU(8)$ induced by
a third rank antisymmetric tensor scalar field}

\author{Stephen L. Adler}
\email{adler@ias.edu} \affiliation{Institute for Advanced Study,
Einstein Drive, Princeton, NJ 08540, USA.}

\begin{abstract}

We study $SU(8)$ symmetry breaking induced by minimizing the Coleman-Weinberg
effective potential for a third rank antisymmetric
tensor scalar field in the 56 representation.  Instead of breaking
$SU(8) \supset SU(3) \times SU(5)$, we find that the stable minimum of the
potential breaks the original symmetry according to $SU(8) \supset SU(3) \times Sp(4)$.
Using both numerical and analytical
methods, we present results for the potential minimum, the corresponding Goldstone
boson structure and BEH mechanism, and the group-theoretic
classification of the residual states after symmetry breaking.

\end{abstract}

\maketitle

\section{Introduction}
In 2014 we proposed a model for family unification \cite{adler1} based on the gauge group $SU(8)$.
The state content of the theory is motivated by the requirement that it should
incorporate the graviton, a set of gravitinos, and the fermions and bosons of the standard
model, with boson-fermion balance but without requiring full supersymmetry. The representation content
of the model is given in Table I, and as shown in \cite{adler1} and was first pointed out by
Marcus \cite{marcus}, $SU(8)$ anomalies cancel between the 8 of gravitinos of the model and the spin $\frac{1}{2}$
fermion fields. Through a detailed analysis we have shown \cite{adler2} that massless spin $\frac{3}{2}$ particles admit a consistent gauging,
both at the classical and quantum levels, so the fact that massive spin $\frac{3}{2}$ cannot be consistently  gauged does not
appear to be an impediment to the anomaly cancellation mechanism of \cite{marcus}, \cite{adler1}.  A further interesting feature of the model is that the $SU(8)$ symmetry forbids
bare Yukawa couplings of the scalar bosons to the fermions, so such couplings can only arise through gauge coupling radiative corrections.    Moreover,  a supersymmetry that is present in the limit of zero gauge
coupling forbids bare quartic couplings of the scalar fields, since these
couplings are not invariant under supersymmetry rotations that mix the representation 56 scalars with representation $\overline{28}$ spin $\frac{1}{2}$ fermions. Thus  if this supersymmetry  is enforced in the model, quartic scalar boson couplings can only arise through radiative corrections.  (Both of these features of the model are discussed in further detail in \cite{adler1}.)
Hence by a criterion proposed by S. Weinberg \cite{sweinberg}, the model is a candidate for a {\it calculable}, as opposed to a merely renormalizable, theory.

\begin{table} [ht]
\caption{Field content of the model of \cite{adler1}.  Square brackets indicate complete antisymmetrization of the enclosed indices.  The indices $\alpha,\beta,\gamma$ range from 1 to 8, and the index $A$ runs from 1 to 63. The top and bottom sections of the table each
contain bosons and fermions satisfying the requirement of boson-fermion balance, with the helicity counts for top and bottom
128 and 112 respectively.}
\centering
\begin{tabular}{ c c c c c}
\hline\hline
field~~~ & spin~~~ & $SU(8)$ rep.~~~ & helicities~~~  \\
\hline
$h_{\mu \nu}$ & 2 & 1 & 2\\
$\psi_{\mu}^{\alpha}$ & Weyl \,3/2 & 8 & 16  \\
$A_{\mu}^A$ & 1 & 63 & 126 \\
$\chi^{[\alpha\beta\gamma]}$&Weyl \, 1/2 & 56 & 112 \\
\hline
$\lambda_{1[\alpha \beta]}$&Weyl\, 1/2&$\overline{28}$ & 56 \\
$\lambda_{2[\alpha \beta]}$&Weyl \,1/2&$\overline{28}$ & 56 \\
$\phi^{[\alpha\beta\gamma]}$& complex\, 0& 56 & 112 \\
\hline\hline
\end{tabular}
\label{fieldcontent}
\end{table}

In order for the model of \cite{adler1} to explain observed physics, the $SU(8)$ symmetry must be broken by a mechanism
employing the only scalar field present, the complex spin 0 field $\phi^{[\alpha\beta\gamma]}$ in the totally antisymmetric
56 representation of $SU(8)$.  Symmetry breaking by rank three antisymmetric tensor fields has been studied by
Cummins and King \cite {cummins1}, Cummins \cite{cummins2}, and Adler \cite{adler3}, assuming an $SU(8)$   invariant fourth degree potential of the form
\begin{equation}\label{fourthorderpot}
V(\phi)=-\frac{1}{2}\mu^2\sum_{\alpha\beta\gamma} \phi^*_{[\alpha\beta\gamma]}\phi^{[\alpha\beta\gamma]}
+\frac{1}{4}\lambda_1\big(\sum_{\alpha\beta\gamma} \phi^*_{[\alpha\beta\gamma]}\phi^{[\alpha\beta\gamma]}\big)^2
+\frac{1}{4}\lambda_2 \sum_{\alpha\beta\gamma\rho\kappa\tau} \phi^*_{[\alpha \beta \gamma]}\phi^{[\alpha\beta\tau]}
\phi^*_{[\rho\kappa\tau]}\phi^{[\rho\kappa\gamma]}~~~.
\end{equation}
When $\mu^2>0$, so that the origin is a local maximum, and  $\lambda_2<0<3\lambda_1+\lambda_2$, which is the simplest case for which
the potential is bounded from below, the original $SU(8)$ symmetry is broken
to $SU(3)\times SU(5)$.  But because bare quartic couplings of the scalar are not allowed by the symmetries of the model, this
potential does not furnish a realistic starting point for studying symmetry breaking in the model of \cite{adler1}.
However, as pointed out in a classic paper of Coleman and E. Weinberg \cite{coleman},  when a massless scalar field is gauged by an
Abelian or non-Abelian gauge symmetry, radiative corrections are the origin of spontaneous symmetry breaking.   Moreover,
the Coleman-Weinberg analysis requires the bare quartic coupling to be sub-dominant if not absent, and so the feature
of the $SU(8)$ model that makes a fourth degree polynomial potential unrealistic is in fact an enabling feature for the Coleman-Weinberg
symmetry breaking mechanism.  With this motivation in mind, the purpose of this paper is to study Coleman-Weinberg
symmetry breaking induced by a gauged third-rank antisymmetric tensor, with a special focus on the case of an $SU(8)$ gauge group.
Contrary to the expectation motivated by the potential of Eq. \eqref{fourthorderpot}, the $SU(8)$ symmetry is not broken to $SU(3) \times SU(5)$, but rather breaks to $SU(3) \times Sp(4)$.

\section{The Coleman-Weinberg effective potential}

We summarize here the results of Sec. VI of Coleman and Weinberg \cite{coleman}, which calculates the one-loop effective potential
for a scalar field in a general massless renormalizable gauge theory.  They sum three types of polygon loops
with scalar field external lines, (i) loops with an internal scalar, (ii) loops with an internal fermion, and
(iii) loops with an internal gauge field.  Because the symmetries of the model of \cite{adler1} forbid scalar
field Yukawa couplings to the fermions, and scalar field quartic couplings in the zero gauge coupling limit,
loops of types (i) and (ii) are absent in this model.   Thus only the gauge field loops contribute, for which
the result of \cite{coleman} (with ${\rm log}$ denoting the natural logarithm) is
\begin{equation}\label{effpot}
V=\frac{3}{64\pi^2} {\rm Tr} M^4(\phi) {\rm log}[M^2(\phi)] ~~~.
\end{equation}
Here $M^2(\phi)$ is the gauge boson mass matrix defined by writing the term in the underlying Lagrangian quadratic
in the gauge potential in the form
\begin{equation}\label{massmatrixdef}
{\cal L}= -\frac{1}{2} \sum_{AB} A_{\mu}^B M^2(\phi)_{BA} A^{\mu \, A}~~~.
\end{equation}
For a rank three antisymmetric scalar interacting with a gauge field with self-adjoint generators $t_A$ and gauge coupling $g$,
one reads off from the Lagrangian of \cite{adler1} the formula for the mass matrix before symmetrization in $A$ and $B$,
\begin{align}\label{massmatrix1}
M^2(\phi)_{BA}=&g^2\big( (t_B)^{\rho^{\prime}}_{\alpha}\delta^{\sigma^{\prime}}_{\beta}\delta^{\tau^{\prime}}_{\gamma}
+(t_B)^{\sigma^{\prime}}_{\beta}\delta^{\rho^{\prime}}_{\alpha}\delta^{\tau^{\prime}}_{\gamma}
+(t_B)^{\tau^{\prime}}_{\gamma}\delta^{\rho^{\prime}}_{\alpha}\delta^{\sigma^{\prime}}_{\beta}\big)\cr
\times& \big((t_A)^{\alpha}_{\rho}\delta^{\beta}_{\sigma}\delta^{\gamma}_{\tau}
+(t_A)^{\beta}_{\sigma}\delta^{\alpha}_{\rho}\delta^{\gamma}_{\tau}
+(t_A)^{\gamma}_{\tau}\delta^{\alpha}_{\rho}\delta^{\beta}_{\sigma}\big) \phi^*_{[\rho^{\prime}\sigma^{\prime}\tau^{\prime}]}
\phi^{[\rho\sigma\tau]}~~~.\cr
\end{align}
After some algebra, this expression can be reduced to the computationally simpler form
\begin{equation}\label{massmatrix2}
M^2(\phi)_{BA}=3g^2 \big( (t_Bt_A)^{\rho^{\prime}}_{\rho} \theta^{\rho}_{\rho^{\prime}} + 2 (t_B)^{\rho^{\prime}}_{\rho} (t_A)^{\tau^{\prime}}_{\tau} \theta^{[\rho \tau]}_{[\rho^{\prime} \tau^{\prime}]}\big)~~~,
\end{equation}
where we have defined
\begin{equation}\label{thetadefs}
\theta^{\rho}_{\rho^{\prime}} \equiv  \phi^*_{[\rho^{\prime}\sigma\tau]}\phi^{[\rho\sigma\tau]}~~,~~~
\theta^{[\rho\tau]}_{[\rho^{\prime}\tau^{\prime}]} \equiv \phi^*_{[\rho^{\prime}\sigma\tau^{\prime}]}\phi^{[\rho\sigma\tau]}~~~.
\end{equation}
As discussed in \cite{coleman}, and also in the text of S. Weinberg \cite{weinbergtext}, the mass matrix that enters into Eq. \eqref{effpot}
is a real symmetric matrix, defined by symmetrizing the expression of Eq. \eqref{massmatrix2} with respect to $A$ and $B$, or equivalently,
taking its real part.  In using Eq. \eqref{massmatrix2} we shall set $g^2$ equal to unity, since it can be absorbed into the implicit
scale mass inside the logarithm in Eq. \eqref{effpot}, and into the overall multiplicative factor which plays no role in determining
the minimum.

Although we will not study rank two antisymmetric tensors in this paper, for completeness we give the corresponding expression
for the mass matrix for a rank two antisymmetric tensor,
\begin{equation}\label{massmaatrix3}
M^2(\phi)_{BA}=2g^2 \big( (t_Bt_A)^{\rho^{\prime}}_{\rho} \theta^{\rho}_{\rho^{\prime}} +  (t_B)^{\rho^{\prime}}_{\rho} (t_A)^{\tau^{\prime}}_{\tau} \theta^{[\rho \tau]}_{[\rho^{\prime} \tau^{\prime}]}\big)~~~,
\end{equation}
where in this case we have defined
\begin{equation}\label{thetadefs1}
\theta^{\rho}_{\rho^{\prime}} \equiv  \phi^*_{[\rho^{\prime}\sigma]}\phi^{[\rho\sigma]}~~,~~~
\theta^{[\rho\tau]}_{[\rho^{\prime}\tau^{\prime}]} \equiv \phi^*_{[\rho^{\prime}\tau^{\prime}]}\phi^{[\rho\tau]}~~~.
\end{equation}
Again, this formula must be symmetrized in $A$ and $B$ to give the mass matrix that enters into Eq. \eqref{effpot}.

\section{Minimum compatible with symmetries of the standard model}

Letting $m_i^2$ with $i=1,...,63$ denote the eigenvalues of the symmetrized $M^2(\phi)$,  and dropping the overall constant
factor in Eq. \eqref{effpot}, the effective potential to be minimized is
\begin{equation}\label{effpot1}
V=\sum_{i=1}^{63} m_i^4 ~ {\rm log}\,m_i^2~~~.
\end{equation}
To calculate $V$ numerically, we used the NAG group program F08FAF (dsyev) to compute the eigenvalues $m_i^2$ of the symmetrized $M^2$. Searching
for the potential minimum was then done using the CERN minimization program MINUIT.
Since  the third rank antisymmetric tensor $\phi$ is complex, this potential is a function of $2\times 56=112$ real parameters,  beyond the capabilities of MINUIT.  A more efficient procedure would be to use a canonical form for
$\phi$ of the type discussed by Cummins \cite{cummins2}, which gives specific results only up to $SU(7)$. An extrapolation
based on Table 2 of \cite{cummins2} suggests that a canonical form for $SU(8)$ will involve over 50 real parameters, simultaneous variation
of which is again beyond the capabilities of MINUIT.

Rather than attempting a search in a very large parameter space, we restricted the problem by searching
for a potential minimum $\overline{\phi}^{[\alpha\beta\gamma]}$   that is compatible with the symmetry group $SU(3)\times SU(2) \times U(1)$ of the standard model.  We started
with an assumed minimum $\overline{\phi}^{[123]}=a$, which corresponds to a symmetry breaking $SU(8)\supset SU(3)\times SU(5)$, found the minimum,
and then looked for stability around this minimum.  We found this minimum was unstable under addition of a second parameter $\overline{\phi}^{[456]}=b$,
corresponding to the symmetry breaking $SU(8) \supset SU(3) \times SU(3) \times SU(2)$, and this minimum was in turn unstable under the addition
of a third parameter $\overline{\phi}^{[178]}=c$, corresponding to the symmetry breaking $SU(8)\supset SU(2) \times SU(3) \times SU(2)$, which
still contains the standard model group.  This potential led to a stable minimum with $a=c$, which  by examining the
unbroken generators (see below) we determined to correspond to breaking the original $SU(8)$
to the group $SU(3) \times Sp(4)$, which contains $SU(2) \times SU(3) \times SU(2)$ as a subgroup.  The values of the parameters $a,\,b,\,c$ at this minimum (from MINUIT followed by a refined search that we coded)  are
\begin{align}\label{paramvalues}
\overline{\phi}^{[123]}=a=&0.59762152222034 ~~~,\cr
\overline{\phi}^{[456]}=b=&0.67199048046134 ~~~,\cr
\overline{\phi}^{[178]}=c=&0.59762152222033 ~~~,\cr
\end{align}
with an uncertainty $O(1)$ in the final digit.  The potential minima corresponding to parametrization of the mass matrix with $a$ only, $a,\,b$ only, and $a,\,b,\,c$ are given in Table II.  Because we have searched only a restricted parameter space, we can only say that these parameters give a local minimum, and cannot guarantee that a lower minimum does not lie a finite distance away in parameter space.

\begin{table} [ht]
\caption{Potential minima obtained by a MINUIT search for the cases $\overline{\phi}^{[123]}=a$ only, $a$ together with $\overline{\phi}^{[456]}=b$,
and $a$ and $b$ together with $\overline{\phi}^{[178]}=c$.}

\centering
\begin{tabular}{ c c }
\hline\hline
parameters ~~~ &~~~ minimum of $V$\\
\hline
$a$ & -3.486115 \\
$a$,\,$b$ & -4.430597 \\
$a$,\,$b$,\, $c$ & -5.346680 \\
\hline\hline
\end{tabular}
\label{potmin}
\end{table}

Since the symmetrized mass matrix is only quadratic in the parameters $a, \, b, \, c$ we explored whether its eigenvalues might
be simple functions of the parameters, and this turned out to be the case.  Of the 63 eigenvalues,  60 are polynomials in  the parameters,
and the remaining 3 are roots of a cubic equation with coefficients that are polynomials in the parameters, as shown in Table III, with
the cubic equation coefficients given by
\begin{align}\label{cubiccoeffs}
B=&-\frac{15}{8}(a^2+b^2+c^2) ~~~,\cr
C=&\frac{1}{4}[9b^2(a^2+c^2)+14a^2c^2]~~~,\cr
D=&-\frac{3}{2}a^2b^2c^2~~~.\cr
\end{align}
Using the potential constructed from the formulas of Table III by solving the cubic equation with the NAG group program C02AKF, a MINUIT
search gave the same minima as given in Table II.

\begin{table} [ht]
\caption{Algebraic formulas for the eigenvalues $m_i^2$, \,i =1,...,63 expressed in terms of  $a, \, b, \, c$, as well as the number
of degenerate eigenvalues of each type.}

\centering
\begin{tabular}{ c c }
\hline\hline
eigenvalue degeneracy  & algebraic formula\\
\hline
  14  &  0 \\
   4  &  $\frac{1}{2}a^2$  \\
   4  &  $\frac{1}{2}c^2$  \\
   4  & $\frac{1}{2}(c-a)^2  $\\
   12 & $\frac{1}{2} (a^2+b^2) $\\
   12 &  $\frac{1}{2} (c^2+b^2)$\\
   6  &  $\frac{1}{2} (a^2+b^2+c^2)$ \\
   4  & $\frac{1}{2}(c+a)^2 $ \\
   1,\,1,\,1  & {3~\rm roots $x_{1,2,3}$ of} ~~$x^3+Bx^2+Cx+D=0$   \\
   \hline\hline

\end{tabular}
\label{abcalgebraic}
\end{table}

This method of constructing $V$ also allowed us to determine that the numerical search showing that
$a=c$ at the minimum gives an exact result.  Letting $a=d-\epsilon,\, c=d+\epsilon$,
then Table III and Eq. \eqref{effpot1} show that near $\epsilon=0$ the potential has the form
\begin{equation}\label{acexpan}
V(\epsilon)=V(0)+32\epsilon^4{\rm log}\epsilon + A\epsilon^2+O(\epsilon^4)~~~.
\end{equation}
Since $\epsilon^2 {\rm log} \epsilon \geq -1/(2e)$, with $e=2.718...$, the derivative $dV/d\epsilon=2\epsilon(A+64\epsilon^2 {\rm log} \epsilon)+O(\epsilon^3)$  vanishes solely at $\epsilon=0$ when
$A> 32/e=11.772$.  A numerical fit to $V$ versus $\epsilon$, with the logarithmic term omitted, shows a parabola near $\epsilon=0$ with
coefficient $A=17.019>11.772$, and so $\epsilon=0$ is a stable potential minimum, that is, $a$ is exactly equal to $c$ at the minimum.
For values of $A$ below $32/e$, $dV/d\epsilon$ has zeros at $\epsilon=\pm \epsilon_*\neq 0$, with $32\epsilon_*^2{\rm log} \epsilon_* \simeq -A/2$.  Substituting into
Eq. \eqref{acexpan}, we see that $V(\pm \epsilon_*) \simeq V(0)+ (A/2) \epsilon_*^2$, so the potential $V$ starts to break the $a=c$ symmetry when $A$ becomes negative.
From Table III, we see that when $a \neq c$, there are 14 zero eigenvalues, corresponding to the generator counting 3+8+3=14 for
the symmetry breaking pattern $SU(8)\supset SU(2) \times SU(3) \times SU(2)$.  However, when $a=c$ there are 18 zero eigenvalues, corresponding
to the generator counting  8+10=18 for the symmetry breaking pattern $SU(8) \supset SU(3) \times Sp(4)$.

When $c=a$, the eigenvalues of Table III simplify to those shown in Table IV, with 61 eigenvalues given as polynomials in $a$ and $b$,
and the remaining two the roots of a quadratic equation with coefficients that are polynomials in the parameters, given by
\begin{align}\label{quadcoeffs}
E=&-\frac{1}{8}(14a^2+15b^2)   ~~~,\cr
F=&\frac{3}{4}a^2b^2  ~~~.\cr
\end{align}
The quadratic polynomial of Table IV is a factor of the cubic polynomial of Table III when it is specialized to  $c=a$ ,
\begin{equation}\label{factor}
x^3-\frac{15}{8}(2a^2+b^2)x^2+\frac{1}{4}(18a^2b^2+14a^4)x-\frac{3}{2}a^4b^2=
[x^2-\frac{1}{8}(14a^2+15b^2)x+\frac{3}{4}a^2b^2](x-2a^2)~~~,
\end{equation}
which yields the correspondence between Tables III and IV.

\begin{table} [ht]
\caption{Algebraic formulas for the eigenvalues $m_i^2$, \,i =1,...,63 when $c=a$, expressed in terms of  $a, \, b$, as well as the number
of degenerate eigenvalues of each type.}

\centering
\begin{tabular}{ c c }
\hline\hline
eigenvalue degeneracy  & algebraic formula\\
\hline
  18  &  0 \\
   8  &  $\frac{1}{2}a^2$  \\\
   24 & $\frac{1}{2} (a^2+b^2) $\\
   6  &  $\frac{1}{2} (2a^2+b^2)$ \\
   5  & $2a^2$ \\
   1,\,1  & {2~\rm roots $x_{1,2}$ of} ~~$x^2+Ex+F=0$   \\
   \hline\hline

\end{tabular}
\label{abalgebraic}
\end{table}

The $SU(3) \times Sp(4)$ form of the unbroken symmetries at the minimum with $c=a$ can be verified explicitly by
examining the form of the 18 generators corresponding to eigenvalue 0.  These all lie in a $SU(7)$ subgroup of
$SU(8)$, with all entries zero in the row and the column labeled by 1.  The 8 $SU(3)$ generators all lie in the $3 \times 3$
submatrix with rows and columns labeled by 4,5,6.  The remaining 10 generators lie in the $4 \times 4$ submatrix
with rows and columns labeled by 2,3,7,8, and are given explicitly by the self-adjoint (but unnormalized) expressions

\begin{align}\label{zerogen}
m_1=&\left(
\begin{array}{cccc}
0&0&0&1\\
0&0&1&0\\
0&1&0&0\\
1&0&0&0\\
\end{array}
\right)~,~~
m_2=\left(
\begin{array}{cccc}
0&0&0&-i\\
0&0&i&0\\
0&-i&0&0\\
i&0&0&0\\
\end{array}
\right)~,~~
m_3=\left(
\begin{array}{cccc}
0&0&1&0\\
0&0&0&-1\\
1&0&0&0\\
0&-1&0&0\\
\end{array}
\right)~,~~
m_4=\left(
\begin{array}{cccc}
0&0&-i&0\\
0&0&0&-i\\
i&0&0&0\\
0&i&0&0\\
\end{array}
\right) \cr
&~~~~~~~~~~~~~~~~~~~~~~~~~~~~\cr
f_1=&\left(
\begin{array}{cccc}
0&1&0&0\\
1&0&0&0\\
0&0&0&0\\
0&0&0&0\\
\end{array}
\right)~,~~
f_2=\left(
\begin{array}{cccc}
0&-i&0&0\\
i&0&0&0\\
0&0&0&0\\
0&0&0&0\\
\end{array}
\right)~,~~
f_3=\left(
\begin{array}{cccc}
1&0&0&0\\
0&-1&0&0\\
0&0&0&0\\
0&0&0&0\\
\end{array}
\right)~,~~\cr
&~~~~~~~~~~~~~~~~~~~~~~\cr
e_1=&\left(
\begin{array}{cccc}
0&0&0&0\\
0&0&0&0\\
0&0&0&1\\
0&0&1&0\\
\end{array}
\right)~,~~
e_2=\left(
\begin{array}{cccc}
0&0&0&0\\
0&0&0&0\\
0&0&0&-i\\
0&0&i&0\\
\end{array}
\right)~,~~
e_3=\left(
\begin{array}{cccc}
0&0&0&0\\
0&0&0&0\\
0&0&1&0\\
0&0&0&-1\\
\end{array}
\right)~.\cr
\end{align}

Under the symmetry breaking $SU(8)\supset SU(3) \times Sp(4)$, the representations 8, $\overline{28}$, 56, and 63 of $SU(8)$ decompose as
follows,
\begin{align}\label{decomp}
8=&(1,4)+(3,1)+(1,1)~~~,\cr
\overline{28}=&(\overline{3},4)+(1,5)+(1,4)+(3,1)+(\overline{3},1)+(1,1)~~~,\cr
56=&(3,5)+(3,4)+(\overline{3},4)+(3,1)+(\overline{3},1)+(1,5)+(1,4)+(1,1)_1+(1,1)_2~~~,\cr
63=&(1,10)+(8,1)+(3,4)+(\overline{3},4)+(3,1)+(\overline{3},1)+(1,5)+(1,4)_1+(1,4)_2+(1,1)_1+(1,1)_2~~~.\cr
\end{align}
For the 63 representation, we can infer from the degeneracy counting the correspondence between the representations
in Eq. \eqref{decomp} and the eigenvalues in Table IV:  $(1,10)+(8,1)$ corresponds to the 18 zero eigenvalues, $(3,4)+(\overline{3},4)$
corresponds to the 24 eigenvalues $\frac{1}{2} (a^2+b^2) $, $(3,1)+(\overline{3},1)$ corresponds to the 6 eigenvalues $\frac{1}{2} (2a^2+b^2)$,
$(1,5)$ corresponds to the 5 eigenvalues $2a^2$, $(1,4)_1+(1,4)_2$ corresponds to the 8 eigenvalues  $\frac{1}{2}a^2$ , and $(1,1)_1+(1,1)_2$
corresponds to the 2 eigenvalues that are the roots of the quadratic equation $x^2+Ex+F=0$ .  Since only the $(1,10)$ and $(8,1)$ generators
in the 63 remain massless after symmetry breaking, the 45 remaining components of the 63,
\begin{equation}\label{higgesed}
(3,4)+(\overline{3},4)+(3,1)+(\overline{3},1)+(1,5)+(1,4)_1+(1,4)_2+(1,1)_1+(1,1)_2
\end{equation}
must pick up scalar partners from those $\phi^{[\alpha\beta\gamma]}$ in the 56 that have the same quantum numbers as in Eq. \eqref{higgesed}.
We shall see in the next section how this comes about.

\section{$SU(3) \times Sp(4)$ and $SU(2) \times SU(3) \times SU(2)$ representation content of $SU(8)$ representations, and identification of the scalars that are absorbed as vector longitudinal components}

We begin by giving in Tables V, VI, VII the representation content of the 8, $\overline{28}$, and 56 representations of $SU(8)$, with respect to the group $SU(3) \times Sp(4)$ that corresponds to the symmetry breaking minimum with $a=c$.  It is simplest to get these by first giving the corresponding representation
content with respect to the group $SU(2) \times SU(3) \times SU(2)$  corresponding to symmetry breaking with $a\neq c$. For this case each antisymmetric tensor
component  corresponds to a unique group representation, which can be read off by inspection of the indices, and
these in turn correspond to one or at most two representations of $SU(3) \times Sp(4)$.  When there are two possible $Sp(4)$ identifications, a
calculation using the generators of Eq. \eqref{zerogen} shows that $[*23]+[*78]$ is in the 1 of $Sp(4)$, and $[*23]-[*78]$ is in the 5 of $Sp(4)$, where $*$ denotes any index not acted on by $Sp(4)$.  (Here, and in the Tables V, VI, VII, we use the abbreviation $\phi^{[\alpha\beta\gamma]}\equiv[\alpha\beta\gamma]$.)

\begin{table} [ht]
\caption{Representation content of $SU(8)$ 8}

\centering
\begin{tabular}{ c c c}
\hline\hline
~~~$SU(2) \times SU(3) \times SU(2)$ ~~~  & ~~~$SU(3) \times Sp(4)$~~~ &~~~ tensor components\\
\hline
 (1,1,1)   & (1,1)  &[1]  \\
(1,3,1)    & (3,1)  &[4],\,[5],\,[6]  \\
 (2,1,1)   & (1,4)  &[2],\,[3]  \\
(1,1,2)    & (1,4)  & [7],\,[8] \\
\hline\hline

\end{tabular}
\label{8cont}
\end{table}

\begin{table} [ht]
\caption{Representation content of $SU(8)$ $\overline{28}$}

\centering
\begin{tabular}{ c c c}
\hline\hline
~~~$SU(2) \times SU(3) \times SU(2)$ ~~~  &~~~$SU(3) \times Sp(4)$~~~ &~~~ tensor components\\
\hline
 (1,1,1)   & (1,1)  & [23]+[78] \\
 (1,3,1)   &(3,1)  & [45],\,[46],\,[56] \\
(1,$\overline{3}$,1)    & ($\overline{3}$,1)  &[14],\,[15],\,[16]  \\
(2,1,1)    &(1,4)   & [12],\,[13] \\
 (1,1,2)   &(1,4)   & [17],\,[18] \\
 (1,1,1)   &(1,5)   &[23]$-$[78]  \\
 (2,1,2)   &(1,5)   & [27],\,[28],\,[37],\,[38] \\
(2,$\overline{3}$,1)    & ($\overline{3}$,4)  & [24],\,[25],[26],\,[34],\,[35],\,[36] \\
(1,$\overline{3}$,2)    & ($\overline{3}$,4)  & [47],\,[48],\,[57],\,[58],\,[67],\,[68]\\
\hline\hline

\end{tabular}
\label{28cont}
\end{table}

\begin{table} [ht]
\caption{Representation content of $SU(8)$ 56}

\centering
\begin{tabular}{ c c c}
\hline\hline
~~~$SU(2) \times SU(3) \times SU(2)$ ~~~  & ~~~$SU(3) \times Sp(4)$~~~&~~~ tensor components\\
\hline
(1,1,1)    & (1,1)  & [123]+[178],\,[456] \\
 (1,1,2)   & (1,4)  &[237],\,[238]  \\
 (2,1,1)   & (1,4)  & [278],\,[378] \\
 (1,1,1)   & (1,5)  & [123]$-$[178] \\
 (2,1,2)   & (1,5)  &[127],\,[128],\,[137],\,[138]  \\
 (1,3,1)   & (3,1)  & [234]+[478],\,[235]+[578],\,[236]+[678] \\
(1,$\overline{3}$,1)    & ($\overline{3}$,1)  &[145],\,[146],\,[156]  \\
(2,3,1)    &(3,4)   & [124],\,[125],\,[126],\,[134],\,[135],\,[136] \\
(1,3,2)    &(3,4)   & [147],\,[148],\,[157],\,[158],\,[167],\,[168] \\
(2,$\overline{3}$,1)    & ($\overline{3}$,4)  &[245],\,[246],\,[256],\,[345],\,[346],\,[356]  \\
 (1,$\overline{3}$,2)   &($\overline{3}$,4)   & [457],\,[458],\,[467],\,[468],\,[567],\,[568] \\
(1,3,1)    & (3,5)    &  [234]$-$[478],\,[235]$-$[578],\,[236]$-$[678]    \\
  (2,3,2)  &  (3,5) &   [247],\,[248],\,[257],\,[258],\,[267],\,[268]  \\
  ~~&~~& [347],\,[348],\,[357],\,[358],\,[367],\,[368]\\
\hline\hline

\end{tabular}
\label{56cont}
\end{table}

To study stability of the potential minimum, and the spectrum of Goldstone bosons that are absorbed in the BEH mechanism,
we expand the potential around the minimum by writing
\begin{equation}\label{expand1}
\phi^{[\alpha\beta\gamma]}=\overline{\phi}^{[\alpha\beta\gamma]}+\sigma^{[\alpha\beta\gamma]}~~~,
\end{equation}
with $\overline{\phi}^{[\alpha\beta\gamma]}$ the potential minimum of Eq. \eqref{paramvalues}, and
$\sigma^{[\alpha\beta\gamma]}$ a small perturbation.  Since $\phi$ and $\sigma$ are complex valued, there are
112 independent perturbation parameters $\sigma_R^{[\alpha\beta\gamma]}$ and $\sigma_I^{[\alpha\beta\gamma]}$ , with
$R$ and $I$ denoting respectively the real and imaginary parts.  We find numerically that the expansion of $V$ around the
minimum has the form
\begin{equation}\label{expand2}
V=V_{min}+\sum_{\alpha\beta\gamma}\sum_{\mu\nu\rho}[ \sigma_R^{[\alpha\beta\gamma]} K^R_{\alpha\beta\gamma|\mu\nu\rho} \sigma_R^{[\mu\nu\rho]}
+\sigma_I^{[\alpha\beta\gamma]} K^I_{\alpha\beta\gamma|\mu\nu\rho} \sigma_I^{[\mu\nu\rho]}+{\rm higher~order~in}~\sigma]~~~,
\end{equation}
with quadratic cross terms between the real and imaginary parts vanishing to within numerical errors.   The numerical results show
that the coefficient matrices $K^R$ and $K^I$ are sparse matrices, which break down into $1\times 1$, $2\times 2$,
and $3 \times 3$ blocks that are easily diagonalized  to find the eigenvectors and eigenvalues.  The resulting
eigenvalues are all greater than or equal to zero, that is, the matrices $K^R$ and $K^I$ are positive semidefinite.  There are 45 zero
eigenvalues, corresponding to the quantum numbers listed in Eq. \eqref{higgesed} of the Goldstone bosons that are absorbed by the
BEH mechanism, with the remaining eigenvalues strictly positive.  Thus the potential extremum of  Eq. \eqref{paramvalues} is a
stable local minimum.

The specific enumeration of eigenvalues of $K^R$ and $K^I$ goes as follows.  We give first results for the $1 \times 1$ blocks, then
for the $2\times 2$ blocks, and finally for the $3 \times 3$ blocks.
\begin{enumerate}
\item The $1\times 1$ blocks.
There are four $1\times 1$ blocks in $K^R$ and seven  $1\times 1$ blocks in $K^I$, all with  eigenvalues $\simeq 0$.  The tensor components and
$SU(3) \times Sp(4)$ state labels for these are summarized in Table VIII, with the abbreviated notation now $\sigma^{[\alpha\beta\gamma]}_{R,I}
\equiv [\alpha\beta\gamma]_{R,I}$.  In addition, there are 12 $1\times 1$ blocks in $K^R$ and 12 $1\times 1$  blocks in $K^I$, all with nonzero
eigenvalue 3.1565.

\begin{table} [ht]
\caption{$1\times 1$ blocks with zero eigenvalues}

\centering
\begin{tabular}{  c c}
\hline\hline
tensor component ~~~ &~~~ $SU(3)\times Sp(4)$ representation\\
\hline
 $[237]_R$     & (1,4) \\
 $[238]_R $    & (1,4) \\
$ [278]_R $     & (1,4) \\
$ [378]_R $     & (1,4) \\
$ [237]_I $     & (1,4) \\
$ [238]_I $     & (1,4) \\
$[278]_I $     & (1,4) \\
$ [378]_I  $    & (1,4) \\
$ [123]_I+[178]_I $& (1,1)\\
$ [456]_I  $    & (1,1)\\
$ [123]_I-[178]_I $& (1,5)\\

\hline\hline

\end{tabular}
\label{11basis}
\end{table}

\item The $2\times 2$ blocks.
There are 12 identical $2\times 2$ blocks in $K^R$, with the matrix structure
\begin{equation}\label{2x2block1}
\left(
\begin{array}{cc}
3.7509&3.3358\\
3.3358&2.9666\\

\end{array}
\right)
\end{equation}
and another 12 blocks in $K^I$ with the matrix structure
\begin{equation}\label{2x2block2}
\left(
\begin{array}{cc}
3.7509&-3.3358\\
-3.3358&2.9666\\

\end{array}
\right)~~~,
\end{equation}
that is, the off-diagonal elements are reversed in sign.  The eigenvalues
of these matrices are $6.6715$ and $\simeq 0$, giving 24 zero eigenvalues in all.   The eigenvectors of $K^R$ and $K^I$
individually are in the $(3,4)+(\overline{3},4)$ representation of $SU(3) \times Sp(4)$,
but there are linear combinations of $K^R$ and $K^I$  eigenvectors that
are pure $(3,4)$ and pure $(\overline{3},4)$.   The tensor components corresponding to
the   $(3,4)+(\overline{3},4)$ states are summarized in Table IX.
\begin{table} [ht]
\caption{Tensor components, for real and imaginary parts, corresponding
to the $SU(3)\times Sp(4)$ $(3,4)+(\overline{3},4)$ states. The listed
components form a basis for the $2 \times 2$ matrices of Eqs. \eqref{2x2block1}
and \eqref{2x2block2}.}

\centering
\begin{tabular}{c c}
\hline\hline
basis element 1~~~ &~~~ basis element 2 \\
\hline
   $[124]$ & $[356]$  \\
   $[125]$ & $[346]$  \\
$[126]$  &  $[345]$  \\
$[134]$  &  $[256]$  \\
$ [135]$  &  $[246]$  \\
$[136] $  &  $[245]$  \\
$[147]$  &  $[568]$  \\
$[148]$  &  $[567]$  \\
$[157]$  &   $[468]$  \\
$[158]$  &  $[467]$  \\
$[167]$   &$[458]$  \\
$[168]$  &$[457] $ \\
\hline\hline
\end{tabular}
\label{24 basis}
\end{table}

In addition, there are two identical $2 \times 2$ blocks in $K^R$ with the matrix structure
\begin{equation}\label{2x2block3}
\left(
\begin{array}{cc}
4.2545&4.2545\\
4.2545&4.2545\\

\end{array}
\right)~~~,
\end{equation}
and another two blocks in $K^I$ with the same form but with the off-diagonal elements reversed in sign.
The eigenvalues of these matrices are 8.5090 and $\simeq 0$, giving 4 zero eigenvalues in all. The corresponding eigenvectors
of $K^R$ and $K^I$ are all in the $(1,5)$ representation of $SU(3)\times Sp(4)$, and the corresponding
tensor components are listed in Table X.

\begin{table} [ht]
\caption{Tensor components, for real and imaginary parts, corresponding
to $(1,5)$  states. The listed
components form a basis for the $2 \times 2$ matrix of Eq. \eqref{2x2block3}
and the similar matrix with diagonal elements reversed in sign.}

\centering
\begin{tabular}{c c}
\hline\hline
basis element 1~~~ &~~~ basis element 2 \\
\hline
   $[127]$ & $[138]$  \\
   $[128]$ & $[137]$  \\
\hline\hline
\end{tabular}
\label{2 basis}
\end{table}

\item The $3\times 3$ blocks.

There are three $3\times 3$ blocks in $K^R$ and $K^I$ with the matrix structure
\begin{equation}\label{3x3block1}
\left(
\begin{array}{ccc}
 1.7703  &0.99533   & 0.99533  \\
 0.99533  & 2.1378  & -1.0186  \\
0.99533   & -1.0186  & 2.1378  \\

\end{array}
\right)~~~,
\end{equation}
apart from reversal in sign of all of the off-diagonal matrix elements  $0.99533$, which does not change
the eigenvalue spectrum.  These matrices have
eigenvalues 3.1565, 2.8893, and $\simeq 0$, giving 6 zero eigenvalues in all.  An analysis of the
eigenvectors shows that 3 of the zero eigenvalues are in the $(3,1)$ representation
of $SU(3)\times Sp(4)$, and 3 are in the $(\overline{3},1)$ representation.   The corresponding tensor
components are listed in Table XI.

\begin{table} [ht]
\caption{Tensor components, for real and imaginary parts, corresponding
to the $(3,1)$ and $(\overline{3},1)$ states. The listed
components form a basis for the $3\times 3$ matrix of Eq. \eqref{3x3block1}
and for similar matrices with  off-diagonal elements $0.99533$ reversed in sign.}

\centering
\begin{tabular}{c c c}
\hline\hline
basis element 1~~~ &~~~ basis element 2&~~~basis element 3 \\
\hline
   $[145]$ & $[236]$ &  [678] \\
   $[146]$ & $[235]$ &  [578] \\
   $[156]$ & $[234]$ &  [478] \\

\hline\hline
\end{tabular}
\label{3 basis}
\end{table}

Finally, there is one $3 \times 3$ block in $K^R$, but not in $K^I$, spanning the basis $[123],\,[178],\,[456]$ corresponding to the quantum
numbers of the potential minimum $\overline{\phi}$.  This block has all positive eigenvalues  36.690, 10.143 and 8.5090.

\end{enumerate}

To conclude, from this enumeration we see that there are exactly 45 zero eigenvalues in the second order perturbation matrix around the minimum, 11 coming
from $1\times 1$ blocks, 24+4=28 from $2\times 2$ blocks, and 6 coming from $3\times 3$ blocks, and these have
quantum numbers corresponding precisely to those of the vector mesons that get masses  by the BEH mechanism as listed in
Eq. \eqref{higgesed}.  In addition, there are 112-45=67 nonzero eigenvalues, 24 coming
from $1\times 1$ blocks, 24+4=28 from $2\times 2$ blocks, and $2\times 6 +3=15$  coming from $3\times 3$ blocks, and one can check that these also
form complete $SU(3)\times Sp(4)$ multiplets.  This finishes the analysis of the bosonic sector that is residual after $SU(8)$ symmetry breaking by the
Coleman-Weinberg potential for a third rank antisymmetric tensor scalar field.  Implications for the fermionic sector of the
model of \cite{adler1} will be taken up elsewhere.

\section{Concluding remarks}

We conclude with three brief remarks.

\begin{enumerate}

\item   We note that there is an unbroken $SU(3)$ group, as needed for color symmetry.  Our analysis suggests that the presence of
an exact color $SU(3)$ group in the standard model may be a hint that symmetry breaking by a rank three antisymmetric tensor
is present at the grand unified theory level, since a three index antisymmetric tensor is a natural $SU(3)$ invariant.

\item  We have verified an exact correspondence between the quantum numbers of the Goldstone modes in the  $56$ that are absorbed as longitudinal
components of the broken symmetry generators, and the quantum numbers of these broken symmetry generators in the $63$.  This should be a
feature of  all cases of the BEH mechanism, but it would be nice to have a general proof.

\item   The unbroken symmetries do not contain a $U(1)$ generator, which will be needed to make
contact with the fermion charge structure of the standard model.  In subsequent work we will study whether this $U(1)$, corresponding to an
additional massless gauge boson, can be generated as an emergent symmetry of the model.

\end{enumerate}

\section{Acknowledgements}

I wish to thank Erick Weinberg for helpful conversations in Aspen giving me a synopsis of his paper with Coleman, and Lee Colbert of
the School of Natural Sciences computer staff for assistance with using the CERN program MINUIT and the NAG routines noted above.
Work on this paper in the summer of 2015  was supported in part by  National Science Foundation
 Grant No. PHYS-1066293 and the hospitality of the Aspen Center for Physics  .


\begin{thebibliography}{99}

\bibitem{adler1}  S. L. Adler, Int. J. Mod. Phys. A {\bf 29}, 1450130 (2014).
\bibitem{marcus} N. Marcus, Phys. Lett.  B {\bf 15l7L}, 383 (1985).
\bibitem{adler2}  S. L. Adler, Phys. Rev. D {\bf 92}, 085022 and 085023 (2015).
\bibitem{sweinberg} S. Weinberg,  Phys. Rev. Lett.  {\bf 29}, 388 (1972).
\bibitem{cummins1} C. J. Cummins and R. C. King, J. Phys. A: Math. Gen. {\bf 17}, 627 (1984).
\bibitem{cummins2} C. J. Cummins, J. Phys. A: Math. Gen. {\bf 19}, 1055 (1986).
\bibitem{adler3}  S. L. Adler, Phys. Lett. B {\bf 744}, 380 (2015).
\bibitem{coleman}  S. Coleman and E. Weinberg, Phys. Rev. D {\bf 7}, 1888 (1973).
\bibitem{weinbergtext} S. Weinberg, {\it The Quantum Theory of Fields, Vol II:  Modern Applications}, Cambridge University Press (1996), p. 298.


\end{thebibliography}
\end{document}